\documentstyle[aps,epsf,multicol]{revtex}
\begin{document}
\draft
\title{Coulomb Blockade in the Fractional Quantum Hall Effect Regime}
\author{Michael R. Geller$^{1,2}$ and Daniel Loss$^3$}
\address{$^1$Institute for Theoretical Physics, University of California, Santa Barbara, California 93106}
\address{$^2$Department of Physics and Astronomy, University of Georgia, Athens, Georgia 30602-2451$^*$}
\address{$^3$Department of Physics and Astronomy, University of Basel, Klingelbergstrasse 82, 4056 Basel, Switzerland}
\date{December 31, 1999}
\maketitle

\begin{abstract}
We use chiral Luttinger liquid theory to study transport through a quantum dot in the fractional quantum Hall effect regime and find rich non-Fermi-liquid tunneling characteristics. In particular, we
predict a remarkable Coulomb-blockade-type energy gap that is quantized in units of the noninteracting level spacing, new power-law tunneling exponents for voltages beyond threshold, and a line shape
as a function of gate voltage that is dramatically different than that for a Fermi liquid. We propose experiments to use these unique spectral properties as a new probe of the fractional quantum Hall 
effect.
\end{abstract}

\vskip 0.10in

\pacs{PACS: 73.23.Hk, 71.10.Pm, 73.40.Hm}       
                     
\begin{multicols}{2}

Despite enormous theoretical and experimental effort during the past decade, the nature of transport in the fractional quantum Hall effect (FQHE) regime of the two-dimensional electron gas
\cite{MacDonald review} remains uncertain. Although chiral Luttinger liquid (CLL) theory \cite{Wen proposal,CLL theory reviews} has successfully predicted transport and spectral properties 
of sharply confined FQHE systems near the center of the $\nu = 1/3$ plateau \cite{CLL experiments}, the situation at other filling factors \cite{other filling factors} and in smooth-edged geometries
\cite{smooth edges} is poorly understood. This has motivated us to consider a new, alternative probe of FQHE edge states. 

In a certain sense, tunneling spectra of single-branch edge states are ultimately measurements of $g$, the dimensionless parameter characterizing a CLL that measures the degree to which it 
deviates from a Fermi liquid, for which $g=1$. In particular, the zero-temperature density-of-states (DOS) of a macroscopic CLL varies as $\epsilon^{{1 \over g} - 1},$ which is responsible for its 
well-known power-law tunneling characteristics. It is not surprising (and will be established below) that transport through a large quantum dot in the FQHE regime is primarily governed by the DOS of 
a {\it mesoscopic} CLL. We shall show here that this finite-size DOS has a remarkable low-energy structure that depends on $g$ in an intricate manner. We therefore propose tunneling through 
a quantum dot in the FQHE regime as a new probe of edge-state dynamics.

It has been appreciated for some time that transport through a strongly correlated FQHE droplet would be interesting in its own right, and this motivated Kinaret {\it et al.} to do their
work on the subject\cite{Kinaret etal}. Their work, which mostly focused on the linear response regime and on small system sizes, led to a number of proposed experiments, which have not been carried
out yet. We would like to emphasize, however, that the experiments proposed by Kinaret {\it et al.}, and by us in the present work, although far from routine, should be possible using current 
nanostructure fabrication techniques. 

The main difference between our work and previous work is that we are the first to directly calculate the retarded electron propagator for a mesoscopic CLL, which has required the development of
finite-size bosonization methods appropriate for the CLL\cite{mesoscopic CLL}. As mentioned, this Green's function has a fascinating low-energy structure, which will be described below. This
result has enabled us to map out a considerable portion of the low-temperature phase diagram for transport through a large quantum dot: In the $\nu = 1/q$ state with $q$ an odd integer we
predict a remarkable Coulomb-blockade-like energy gap of size $(q-1) \Delta \epsilon,$ where $\Delta \epsilon$ is the noninteracting level spacing. Unlike a conventional Coulomb blockade \cite{coulomb
blockade}, however, the energy gap here is precisely quantized. Furthermore, the low-temperature tunneling current scales nonlinearly with voltage as $V^q$ at a Coulomb blockade tunneling peak, as
one might expect, but as the voltage is increased between these peaks to overcome the Coulomb blockade the current at the threshold varies as $V^{q+1}.$ The finite-bias line shape as a function of 
gate voltage depends nontrivially on $q$ and is also dramatically different than that for a Fermi liquid.

The model we adopt here for the quantum dot system is as follows: Two macroscopic $g=1$ edge states, L and R, are weakly coupled to a mesoscopic FQHE edge state, D, in the quantum dot, by a
tunneling perturbation
\begin{equation}
\delta H = \sum_{I=L,R} \gamma_{\rm I} \,  \psi_{\rm I}(x_{\rm I}) \, \psi_{\rm D}^\dagger(x_{\rm I})  + \gamma_{\rm I}^* \, \psi_{\rm D}(x_{\rm I}) \, \psi_{\rm I}^\dagger(x_{\rm I}).
\end{equation}  
The edges of the two-dimensional electron gas are assumed to be sharply confined, and the interaction short-ranged (screened by a nearby gate), so that the low lying  excitations consist of a single 
branch of edge-magnetoplasmons with linear dispersion $\omega = v |k|$. Although we will be working at zero-temperature, it is assumed that there is a small temperature present to help suppress
coherence and resonant tunneling. Finally, the dot should be large enough (a few $\mu m$ in circumference) so that the charging energy is smaller than the bulk FQHE energy gap.

What property of the electron gas is probed in a measurement of tunneling through the dot? It is known that the conductance of a simple resistive barrier measures the transmission probability of that 
barrier, a one-particle property, along with the single-particle Green's function of the leads, even when there is strong electron-electron interaction\cite{Schrieffer etal}. In contrast, transport 
through a quantum dot containing other electrons generally probes two-particle (and higher order) properties of the dot, even if interactions in the leads are ignored, because the dot itself has its 
own internal dynamics\cite{one-particle footnote}. However, it is clear that for a large, weakly coupled dot, and small enough currents, an electron can tunnel onto the dot, dissipate energy, and 
then tunnel incoherently through the second barrier, and in this so-called sequential tunneling limit the resistance will probe the one-particle Green's function of the dot. 

To establish this relationship we write the current from $L$ to $R$ as
\begin{equation}
I = \sum_N P(N) \big[ w_{\rm LD}(N) - w_{\rm DL}(N) \big],
\end{equation}
where $w_{\rm LD}$ and  $w_{\rm DL}$ are transition rates to go from $L$ to $D$ and from $D$ to $L$, given that there are $N$ electrons in the quantum dot, and $P(N)$ is the probability that the 
dot has $N$ electrons. The zero-temperature rate from an initial ground state $|\Psi_0^{N_{\rm L}} \rangle_{\rm L} \otimes |\Psi_0^{N} \rangle_{\rm D}$ to final excited states of the form 
$|\Psi_{\alpha_{\rm L}}^{N_{\rm L}-1} \rangle_{\rm L} \otimes |\Psi_{\alpha_{\rm D}}^{N+1} \rangle_{\rm D}$ is given by
\end{multicols}
\begin{equation}
w_{\rm {LD}}(N) = 2 \pi |\gamma_{\rm L}|^2 \sum_{\alpha_{\rm L}} \big| \big\langle \Psi_{\alpha_{\rm L}}^{N_{\rm L}-1} \big| \psi \big| \Psi_{0}^{N_{\rm L}} \big\rangle \big|^2 \sum_{\alpha_{\rm D}} 
\big| \big\langle \Psi_{\alpha_{\rm D}}^{N+1} \big| \psi^\dagger \big| \Psi_{0}^{N} \big\rangle \big|^2 \, \delta \big( E_{\alpha_{\rm L}}^{N_{\rm L}-1} - E_0^{N_{\rm L}} + E_{\alpha_{\rm D}}^{N+1}
 - E_0^{N} - V_{\rm L} +  V_{\rm D} \big),
\end{equation}                                         
where $V_{\rm L}$ and $V_{\rm D}$ are potential energies produced by gates above $L$ and $D$. This can be written as
\begin{equation}
w_{\rm {LD}}(N) = 2 \pi |\gamma_{\rm L}|^2 \ \Theta(V) \ \int_{0}^V d\epsilon \ A_+^{\rm D}(\epsilon) \ A_{-}^{\rm L}( V  - \epsilon),
\end{equation}
where $V$ is the electrochemical potential difference between $L$ and $D$, $\Theta$ is the unit step function, and
\begin{eqnarray}
A_{+}(\omega) &\equiv& \sum_\alpha \big|\big\langle \Psi_\alpha^{N+1} \big| \psi^\dagger \big|\Psi_0^N \big\rangle\big|^2 \, \delta \big(\omega + \mu^N - E_\alpha^{N+1} + E_0^N \big) \\
A_{-}(\omega) &\equiv& \sum_\alpha \big|\big\langle \Psi_\alpha^{N-1} \big| \psi \big|\Psi_0^N \big\rangle\big|^2 \, \delta \big(\omega - \mu^{N-1} - E_\alpha^{N-1} + E_0^N \big),
\end{eqnarray}
where $\mu^N \equiv E_0^{N+1} - E_0^N.$ The chemical potential in the quantum dot is $N$ dependent. With noninteracting leads,
\begin{equation}
w_{\rm {LD}}(N) = 2 \pi |\gamma_{\rm L}|^2  N_{\rm L}(0) \ \Theta(V)  \int_{0}^V d\epsilon \ A_+^{\rm D}(\epsilon),
\label{transition rate}
\end{equation}
where $N_{\rm L}(0)$ is the density of states at the Fermi energy in the left lead. If we define the interacting DOS as $N(\epsilon) \equiv - {1 \over \pi} {\rm Im} \, G(0,\epsilon)$,
where $ G(x,t) \equiv -i \Theta (t) \langle \lbrace \psi_\pm(x,t) , \psi_\pm^\dagger(0) \rbrace \rangle $ is the retarded electron propagator calculated in the grand-canonical ensemble with chemical 
potential $\mu$, then for $\epsilon > 0$ it follows that $A_+(\epsilon) = N(\epsilon)\big|_{\mu = \mu^N}$.

\begin{figure}
\epsfbox{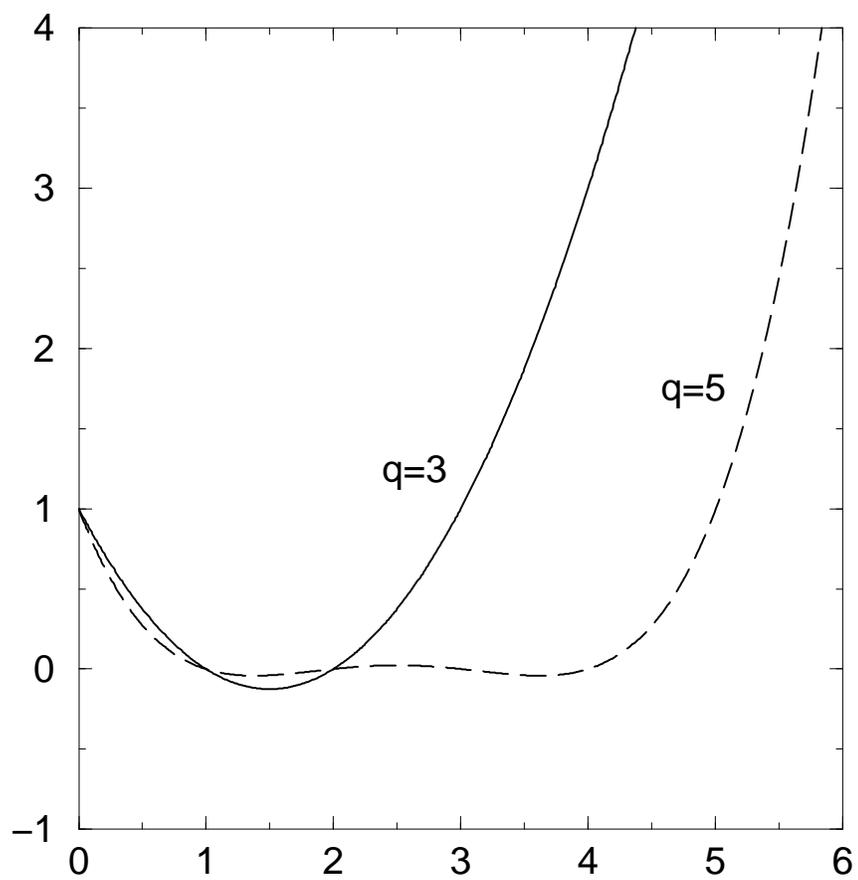}
\caption{Polynomial factor for the cases $q=3$ and 5, plotted as a function of $\omega/\Delta \epsilon$.}
\label{DOS factor}
\end{figure}

The dynamics of the mesoscopic CLL is governed by the action ($g=1/q$ with $q$ an odd integer)
\begin{equation}
S = {1 \over 4 \pi g} \int_0^L \! \! dx \int_0^\beta \! \! d\tau \big[ \pm i (\partial_\tau \phi_\pm)(\partial_x \phi_\pm) + v (\partial_x \phi_\pm)^2 \big] ,
\label{action}
\end{equation}
where $\rho_{\pm} = \pm \partial_x \phi_{\pm} / 2\pi$ is the charge density fluctuation for right (+) or left (--) moving electrons\cite{CLL theory reviews}. Momentum space quantization is achieved 
by decomposing the chiral scalar field $\phi_{\pm}$ into a nonzero-mode contribution $\phi_\pm^{\rm p}$ satisfying periodic boundary conditions, and a zero-mode part $\phi_{\pm}^0$. The bosonized 
electron field is $\psi_{\pm}(x) \equiv (2\pi a)^{-{1\over 2}} e^{iq\phi_\pm(x)} e^{\pm iq\pi x/L}$, where $a$ is a microscopic cutoff length. In the presence of an Aharonov-Bohm flux $\Phi = 
\varphi \ \! \Phi_0$  (with $\Phi_0 \equiv hc/e$) the grand-canonical Hamiltonian corresponding to (\ref{action}) is 
\begin{equation}
H = {1 \over 2g} \, \big(N \pm g \varphi)^2 \Delta \epsilon + \sum_k \Theta(\pm k) \, v |k| \, a_k^\dagger a_k - \mu N,
\label{hamiltonian}
\end{equation}
where $\Delta \epsilon \equiv 2 \pi v /L$ is the noninteracting level spacing and $N \equiv \int_0^L dx \ \rho_{\pm}$. At zero temperature,
$$
G(x,t) = {\textstyle{-i \over 2 \pi a}} \ \Theta(t) \, e^{\pm i q \pi (x \mp vt)/L} \big\langle e^{iq( \phi_\pm^0(x,t) - \phi_\pm^0(0) )} \big\rangle
\big( e^{{1 \over 2} q^2[ \phi_\pm^0(x,t),\phi_\pm^0(0) ]} \ e^{q^2 f_\pm(x,t)} + e^{-{1 \over 2} q^2[ \phi_\pm^0(x,t),\phi_\pm^0(0) ]} \ e^{q^2 f_\pm(-x,-t)} \big),
$$               
where $f_\pm(x,t) \equiv \langle \phi_\pm^{\rm p}(x,t) \phi_\pm^{\rm p}(0) - ( \phi_\pm^{\rm p}(0) )^2 \rangle.$ The time-evolution of the zero-mode field under the action of (\ref{hamiltonian}) is
found to be $\phi_{\pm}^0(x,t) = \pm 2 \pi N (x \mp vt)/L - g \, \chi + g (\mu \mp \varphi \Delta \epsilon)t, $ where $[\chi, N]=i.$ Then
\begin{equation}
G(x,t) = \pm \, \Theta(t) \, (i/L)^q \, (\pi a)^{q-1} \, e^{\pm i q \pi (x \mp vt)/L} \, e^{i(\mu \mp \varphi \Delta \epsilon)t} \,  e^{\pm 2 \pi i q \langle N \rangle (x \mp vt)/L} \, {\rm Im} \, 
\sin^{-q}[\pi(x \mp vt \pm ia)/L],
\label{zero temperature G}
\end{equation}
where $\langle N \rangle = q^{-1} \, {\rm int} ({\textstyle{ \mu \over \Delta \epsilon}} \mp \varphi)$. Here ${\rm int}(x)$ denotes the integer closest to $x$. The transform may be written as
\begin{equation}
G(x,\omega) = - {\textstyle{i \over \pi v}} ({\textstyle{i \pi a \over L}})^{q-1} \, e^{\pm 2 \pi i q (\langle N \rangle + {1 \over 2})x/L} \int_0^\infty dt \ e^{i \Omega t} \ {\rm Im} \,
\bigg[{1 \over \sin^q (t \mp {\pi x \over L} - i {\pi a \over L}) } \bigg],
\label{G transform}
\end{equation}
where $\Omega \equiv 2[{\omega \over \Delta \epsilon} -q( \langle N \rangle + {1 \over 2}) + {\mu \over \Delta \epsilon} \mp \varphi].$ Note that $\Omega$ depends on $q$ both explicitly and implicitly
through $\langle N \rangle$. To evaluate (\ref{G transform}) we need integrals of the form $\int_0^\infty dt \ e^{i \Omega t} \ {\rm Im} \sin^{-q}(t + \xi -i\eta)$, which
we evaluate by using the identity
\begin{equation}
\int_0^{2 \pi} dt \ e^{i \Omega t} \ {\rm Im} \sin^{-q}(t + \xi -i \eta) = { (q-2)^2 - \Omega^2 \over (q-1)(q-2) } \cdot \int_0^{2 \pi} dt \ e^{i \Omega t} \ {\rm Im} \sin^{-(q-2)}(t + \xi -i \eta).
\ \ \ \ \ \ \ \ q > 2 
\end{equation}
After considerable manipulation we obtain
\begin{equation}
G(x,\omega) = { (i \pi a / L)^{q-1} \over (q-1)!} \, (1-\Omega^2)(3^2 - \Omega^2) \times \cdots \times [(q-2)^2 - \Omega^2]  \times G_0(x,\omega),
\end{equation}
where $G_0(x,\omega)$ is the retarded propagator for the noninteracting chiral electron gas, given below. The $q$ dependence of $\Omega$ is extracted by writing $\Omega = 2 z - q$, 
where $z \equiv {\omega \over \Delta \epsilon} + {\rm frac} ({\mu \over \Delta \epsilon} \mp \varphi)$ and ${\rm frac}(x) \equiv x - {\rm int}(x)$. Finally, after using the identity (proved by 
induction) for $q$ an odd integer greater than one,
\begin{eqnarray}
(1-\Omega^2)(3^2 - \Omega^2) \times \cdots \times [(q-2)^2 - \Omega^2] &=& [1-(2 z -q)^2][3^2 - (2z -q)^2] \times \cdots \times [(q-2)^2 - (2 z -q)^2] \nonumber \\
&=& (2i)^{q-1} \prod_{j=1}^{q-1}(z -j),
\end{eqnarray}
we arrive at the remarkable relation
\begin{equation}
G(x,\omega) = G_0(x,\omega) \times {1 \over (q-1)! \, \epsilon_{\rm F}^{q-1}} \prod_{j=1}^{q-1} \big(\omega - \omega_j \big) ,
\label{remarkable relation}
\end{equation}
where $\epsilon_{\rm F} \equiv v/a$ is an effective Fermi energy and where $\omega_j \equiv [j - {\rm frac}({\mu \over \Delta \epsilon} \mp \varphi)] \, \Delta \epsilon$ are the noninteracting
energy levels. Whereas in the $q=1$ case the propagator has poles at each of the $\omega_j$, in the interacting case the first $q-1$ poles above $\mu$ are {\it removed}. This effect, which leads to
a Coulomb-blockade-type energy gap, is a consequence of the first term in the Hamiltonian (\ref{hamiltonian}). At higher 
frequencies or in the large $L$ limit where $\omega \gg \Delta \epsilon,$ the additional factor becomes $\omega^{q-1} / (q-1)! \, \epsilon_{\rm F}^{q-1}$.
The polynomial factor in Eqn.~(\ref{remarkable relation}) is plotted in Fig.~\ref{DOS factor}.

In (\ref{hamiltonian}) we have taken the single-particle dispersion to be $\epsilon_\pm(k) = \pm v(k + 2 \pi \varphi/L)$. The noninteracting chiral propagator is therefore 
$G_0(0,\omega) = (1/2v) \cot [\theta(\omega)/2],$ where $\theta(\epsilon) = 2 \pi (\epsilon/\Delta \epsilon \mp \varphi)$ is the phase subjected to an electron of energy $\epsilon$ 
after going around the edge state.

Having obtained the transition rate (\ref{transition rate}) we turn to a calculation of the probability $P(N)$, which satisfies
\begin{equation}
\partial_t P(N) = \sum_{I=L,R} \big[ w_{\rm ID}(N-1) \ P(N-1) + w_{\rm DI}(N+1) \ P(N+1) - w_{\rm ID}(N) \, P(N) - w_{\rm DI}(N) \, P(N) \big].
\label{master equation}
\end{equation}
The steady-state solution of (\ref{master equation}) yields the final result for the tunneling current. For the case $q=3$,
\begin{equation}
I = 2 \pi |\gamma|^2 \, [N(0)]^2 \, { V^2(V - {4 U^2 \over V} [N_{\rm G} - (n+{1 \over 2})]^2 )^3 \over V^2 + 12 U^2 [N_{\rm G}-(n+{1 \over 2})]^2} \ \ \ \ \ {\rm when} \ \ V > 2 U |N_{\rm G} - 
(n+{\textstyle{1 \over 2}})|,
\label{final current}
\end{equation}
and is zero otherwise. Eqn.~(\ref{final current}) is valid for $n < N_{\rm G} < n+1$, where the gate charge $N_{\rm G}$ is the number of positive 
charges induced by the gate, and for symmetric leads. $U \equiv q \, \Delta \epsilon$ is the quantized charging energy plus the single-particle level spacing.

\begin{multicols}{2}

\begin{figure}
\begin{center}
\epsfbox{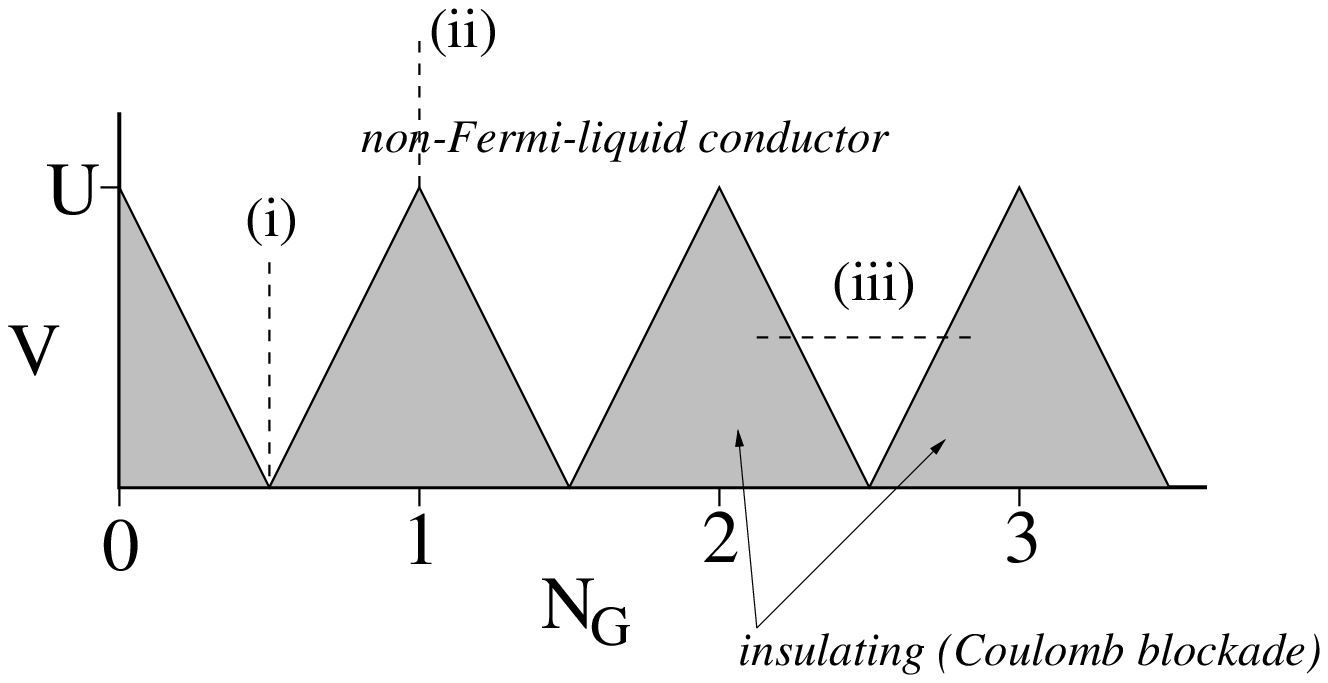}
\end{center}
\caption{Phase diagram for the tunneling current as a function of bias voltage $V$ and gate charge $N_{\rm G}$.}
\label{phase diagram}
\end{figure}

The $q=3$ result (\ref{final current}) clearly exhibits the novel transport properties present at all $q \neq 1$. The Coulomb blockade boundary, shown as a solid line in Fig.~\ref{phase diagram}, has
the familiar diamond shape, but the scale $U$ is now quantized in units of $\Delta \epsilon$. It can be shown that the current in a Fermi liquid would be proportional to a term of the form
$V-{4 U^2 \over V}[N_{\rm G}-(n+{1 \over 2})]^2$ alone. The additional structure present in Eqn.~(\ref{final current}) describes how the quantum dot becomes a non-Fermi-liquid conductor when 
threshold is exceeded. Examples of this non-Fermi-liquid behavior are shown in Fig.~\ref{phase diagram}. Along path (i) the current varies as $V^q$, as one might naively expect, but on (ii) it varies 
as $(V-U)^{q+1}$. The line shape along (iii) depends nontrivially on $q$; for $q=3$ it varies as $(1-4x^2)^3/(1+12 x^2)$, which, surprisingly, is in excellent agreement with the finite-bias numerical 
results for just 8 electrons\cite{Kinaret etal}. The transport properties at other values of $q$ can be determined from Eqn.~(\ref{remarkable relation}).
             
This work was supported by the National Science Foundation under Grant No.~PHY94-07194, by a Research Innovation Award from the Research Corporation, and by the Swiss National Science Foundation. 
It is a pleasure to thank Matt Grayson and Marc Kastner for useful discussions. 

\end{multicols}

\begin{references}

\bibitem{MacDonald review} For a recent review, see A. H. MacDonald in {\it Mesoscopic Quantum Physics}, edited by E. Akkermans {\it et al.} (Elsevier, Amsterdam, 1995). 

\bibitem{Wen proposal} X. G. Wen, Phys. Rev. B {\bf 43}, 11025 (1991).

\bibitem{CLL theory reviews} For reviews, see X. G. Wen, Adv. Phys. {\bf 44}, 405 (1995), and C. L. Kane and M. P. A. Fisher, in {\it Perspectives in Quantum Hall Effects: Novel Quantum Liquids in
Low-Dimensional Semiconductor Structures}, edited by S. Das Sarma and A. Pinczuk (John Wiley, New York, 1997).

\bibitem{CLL experiments} F. P. Milliken {\it et al.}, Solid State Commun. {\bf 97}, 309 (1996). A. M. Chang {\it et al.}, Phys. Rev. Lett. {\bf 77}, 2538 (1996). L. Saminadayar {\it et al.}, Phys. 
Rev. Lett. {\bf 79}, 2526 (1997). R. de-Picciotto {\it et al.}, Nature {\bf 389}, 162 (1997). 
                                            
\bibitem{other filling factors} M. Grayson {\it et al.}, Phys. Rev. Lett. {\bf 80}, 1062 (1998).        

\bibitem{smooth edges}  J. D. F. Franklin {\it et al.}, Surf. Sci. {\bf 361}, 17 (1996). I. J. Maasilta and V. J. Goldman, Phys. Rev. B {\bf 55}, 4081 (1997).

\bibitem{Kinaret etal} J. M. Kinaret {\it et al.}, Phys. Rev. B {\bf 45}, 9489 (1992). J. M. Kinaret {\it et al.}, Phys. Rev. B {\bf 46}, 4681 (1992).

\bibitem{mesoscopic CLL} F. D. M. Haldane, J. Phys. C {\bf 14}, 2585 (1981). X. G. Wen, Phys. Rev. B {\bf 41}, 12838 (1990). M. Stone, Ann. Phys. {\bf 207}, 38 (1991). M. R. Geller and D. Loss, Phys.
Rev. B {\bf 56}, 9692 (1997).

\bibitem{coulomb blockade} D. V. Averin and K. K. Likharev, in {\it Mesoscopic Phenomena in Solids}, edited by B. L. Altshuler {\it et al.} (Elsevier, New York, 1991).
                                                    
\bibitem{Schrieffer etal} J. R. Schrieffer, D. J. Scalapino, and J. W. Wilkins, Phys. Rev. Lett. {\bf 10}, 336 (1963).

\bibitem{one-particle footnote} The current, being the expectation value of a one-particle operator, can be expressed in terms a one-particle Green's function [see Y. Meir and N. S. Wingreen, Phys. 
Rev. Lett. {\bf 68}, 2512 (1992)], but in this case the Green's function is no longer that of the quantum dot itself.
 


\end{references}
\end{document}